# Title: A framework for quantitative analysis of Computed Tomography images of viral pneumonitis: radiomic features in COVID and non-Covid patients

**Authors:** G. Zorzi[1], L. Berta[2], S. Carrazza[1,3], A. Torresin[1,2]

[1] Department of Physics, Università degli Studi di Milano, via Giovanni Celoria 16, 20133 Milan, Italy;

[2] Department of Medical Physics, ASST Grande Ospedale Metropolitano Niguarda, Piazza Ospedale Maggiore 3, 20162 Milan, Italy;

[3] Department of Physics, INFN Sezione di Milano, via Giovanni Celoria 16, 20133 Milan, Italy;

**Corresponding author:** giulia.zorzi1@studenti.unimi.it

**Abstract**

**Purpose**: to optimize a pipeline of clinical data gathering and CT images processing implemented during the COVID-19 pandemic crisis and to develop artificial intelligence model for different of viral pneumonia.

**Methods**: 1028 chest CT image of patients with positive swab were segmented automatically for lung extraction. A Gaussian model developed in Python language was applied to calculate quantitative metrics (QM) describing well-aerated and ill portions of the lungs from the histogram distribution of lung CT numbers in both lungs of each image and in four geometrical subdivision. Furthermore, radiomic features (RF) of first and second order were extracted from bilateral lungs using PyRadiomic tools. QM and RF were used to develop 4 different Multi-Layer Perceptron (MLP) classifier to discriminate images of patients with COVID (n=646) and non-COVID (n=382) viral pnumonia.

**Results**:

The Gaussian model applied to lung CT histogram correctly described healthy parenchyma 94% of the patients. The resulting accuracy of the models for COVID diagnosis were in the range 0.76-0.87, as the integral of the receiver operating curve. The best diagnostic performances were associated to the model based on RF of first and second order, with 21 relevant features after LASSO regression and an accuracy of 0.81±0.02 after 4-fold cross validation

**Conclusions**: Despite these results were obtained with CT images from a single center, a platform for extracting useful quantitative metrics from CT images was developed and optimized. Four artificial intelligence-based models for classifying patients with COVID and non-COVID viral pneumonia were developed and compared showing overall good diagnostic performances

1. Introduction

COVID-19 is an acute respiratory illness caused by a coronavirus. The most common symptoms are fever, fatigue, loss of taste or smell, cough and shortness of breath but can also become much more severe including acute respiratory distress syndrome (ARDS) and requiring the help of external oxygen or invasive ventilation and in some cases it can even lead to death; especially in elderly people or those who are in poor health conditions. The Berlin definition of ARDS also provides for a classification of patients based on oxygenation: according to a clinical study conducted in New York of over 5200 patients, the average mortality for patients with a critical form of COVID-19 is 49%.

The virus is transmitted directly between people through particles emitted by air. For the first time it was identified as a cluster of pneumonia in Wuhan, China, in 2019 and then it spread rapidly and became pandemic in 2020. The pandemic was officially announced by theWorld Health Organization on March 11, 2020. The name COVID-19 was assigned by the World Health Organization and refers to "Coronavirus disease 2019". To date, the cases ascertained all over the world are more than 224 millions and in Italy they are 4,62 millions. Chest Computer Thomography (CT) plays an important role in the diagnosis, detection of pathology and prognosis of COVID-19; it can characterize lung involvement recognizing different imaging patterns based on the tissue density. The most frequent radiographic findings are ground glass consolidations and opacities, with bilateral, peripheral and lower lung lobe distributions. Most COVID-19 studies use a qualitative approach, describing lesions via visual assessment. In the first months of pandemic emergency, the Netherlands Radiological Society has developed the *Coronavirus disease 2019 Reporting and Data System* (CO-RADS) in order to standardize terminology and communication and evaluate a possible presence of COVID-19. It is a categorical scheme that evaluates the pulmonary involvement of the disease observed with non-contrast thoracic CT. In addition, a quantitative approach, visual quantitative analysis, was also needed to evaluate the effects of COVID-19 on the lungs. The latter includes a score of the pulmonary anomalies assessed by visual interpretation of the CT images and densitometric evaluations based on the histogram analysis of the distribution of the Hounsfield Unit (HU). This approach has proved to be useful in predicting the clinical severity of the disease. Along with medical imaging, artificial intelligence (AI) coupled with machine learning technology has achieved impressive results and can be exploited in this scenario.

The aim of this study was to build a pipeline for the automatic extraction of *biomarkers* through the quantitative analysis of the images. In the field of medical physics the word biomarkers refers to the statistical values obtained from quantitative analysis because they are measurable values that can be correlated to clinical aspects.

Exploiting the extracted features, a classifier, capable of predicting the COVID-19 illness is trained and validated.

## 2. Material and Methods

This retrospective study was approved by the Local Ethics Committee. The need for informed consent was waived owing to the retrospective nature of the study. This section is divided in 3 parts…

### 2.1. Dataset

The image dataset is composed by 1028 chest DICOM CTs of patients with viral pneumonia. The average age of the sample is 65.3 years, in range 18-99 years and the percentage of male patients is higher than that of females (67% compared to 33%). The population consists of 646 COVID-19 and 382 non COVID-19. COVID-19 CT scans refer to patients with a positive swab for COVID-19 and are divided into:

• 60 patients from the first wave (February 2020 - June 2020);

• 152 patients from the second wave (September 2020 - December 2020);

• 434 patients from the third wave (January 2021 - April 2021).

This group of patients is made up of 191 women and 455 men in the age group 18-99 years.

Non-COVID-19 patients have a positive swab for H1N1 viral pneumonia in the 2015-2019 period. 147 patients are female and 235 are male; they are aged between 19 and 98. This patient cohort was chosen from among patients with H1N1 as this virus causes pneumonia showing signs in the lungs similar to those produced by COVID-19.

### 2.2. Lung Segmentation

Chest CT dataset DICOM images, both COVID-19 and no COVID-19, were converted to NIfTI format using ImageJ software. Without applying any kind of preprocessing, the latter were automatically segmented with the use of the *lungmask* (version 0.2.9) package by exploiting the U-net (R231) convolutional network trained on COVID patients (R231CovidWeb). This software automatically distinguishes the left and right lungs.

The masks were processed in JavaScript to form two types of lung masks called "bilateral" and "Sub-ROI".

The bilateral mask was obtained by combining the left and right masks. The Sub-ROI mask was created considering a ventral-dorsal and superior-inferior subdivision. The upper-lower subdivision was obtained so that the two portions had the same volume. Then, in each axial section of the upper lung region, the line connecting the left and right lung centroids was used to separate the ventral and dorsal regions. The ventral-dorsal subdivision was then calculated for the lower region by extending the results of the lower slice into the upper lung region. Finally, the left and right portions of the lungs were joined.

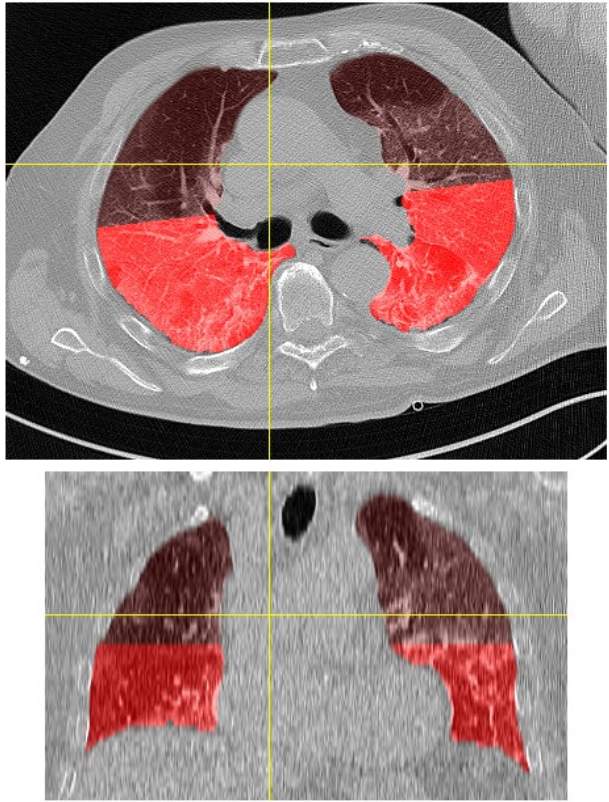

Figure 1: Lung segmentation with subdivision of the mask into ventral-dorsal and superior-inferior.

### 2.3. Quantititve metrics and Features extraction

The features necessary for the construction of artificial intelligence models have been extracted in two different ways: quantitative analysis of the image carried out with an in-house code and using PyRadiomics package.

#### 2.3.1. Quantitative metrics from CT histogram (QCT)

A software was developed in Python3 language to quantitatively analyze the distribution of voxel values in lung CT images. It requires as input a 3D thoracic acquisition and a mask produced by a segmentation. Then, it applies the mask to the lung and perform an analysis concerning only the voxels identified by the lung mask. The method estimates the well aerated volume assuming that the distribution of the HU values of the voxels concerning exclusively healthy parenchymal tissue are described by a Gaussian function (Gaussian model). It allows to perform a histogram analysis of the HU distribution and to extract a series of biomarkers such as mean, standard deviation, asymmetry, kurtosis and percentiles. Then starting from the hypothesis that the well aerated portion of the lung is represented by a Gaussian-shaped peak, the aerated region and the region defined as diseased (consisting of consolidations and vessels) are identified: the well aerated part of the lung is identified with a Gaussian fit, while the diseased part of the lung is obtained by subtracting the gaussian function from the histogram of the HU distribution. After identifying the two regions, a series of biomarkers of the fit (i.e. WAVE.f) and biomarkers of the diseased parts (i.e. mean.ill, standard deviation.ill, skewness.ill, kurtosis.ill, percentiles.ill) are extracted. The software allows the analysis and extraction of biomarkers in bilateral lung and in four sub-regions with two separate codes.

Starting from what has already been developed, in this work the Gaussian model was extended by simultaneously analyzing the bilateral lung and four sub-regions: ventral-dorsal and inferior-superior. In this way, the biomarkers are extracted simultaneously in all the regions considered and offer information
on the localization of the disease. In the following we will refer to the four subregions and the bilateral lung with the term "five regions". The simultaneous analysis of the five regions made it possible to improve the quantitative analysis allowing the extraction of biomarkers even in patients and/or regions where previously, due to the applicability conditions of the Gaussian model, it was not possible. This quantitative analysis was then used to extract biomarkers from bilateral lung and sub-regions; the "bilateral" mask was used, previously rescaling with ImageJ all the images and masks to a thickness of 3 *mm*. The volume associated with the voxels, assuming to have an average value of 0.73 *mm/pix* and 3 *mm* of thickness, is equal to 1.6 $mm^3$. The classifiers called *clinical models* were built using the biomarkers extracted
with this analysis.

### 2.3.2. PyRadiomic

In parallel with the QCT analysis for biomarker extractions, other radiomic features were extracted using PyRadiomics (v3.1), an open source Python package. In this case the features were extracted from the CT images and masks resampled in isotropic voxels of 1.15 mm using ImageJ software. The choice of this isotropic voxel is linked to the fact that a volume of 1.52 $mm^3$ is obtained; this value is similar to that obtained in the case of the non-isotropic voxel used for the QCT analysis which is equal to 1.6 $mm^3$. The extracted features are: the first-order features, those of gray level cooccurence matrix and those of gray level size zone matrix.

A range of HU equal to -1020,180 was used. For the features of the first order, due to the stability as the bin width varies, we have set a bin width equal to 5 and therefore a number of bins of 240.

The GLCM and GLSZM features, on the other hand, have been extracted with different bin width values:

• GLCM: bin width = 5, 25, 50 and numbers of bin = 240, 48, 24;

• GLSZM: bin width = 25, 100, 200 numbers of bin = 48, 12, 6.

They were then used for the construction of the models called radiomic models.

### 2.4. Classifiers

Once the biomarkers were extracted from the images, we moved on to the modelling of COVID classifiers. The goal was to discriminate a COVID population from a non-COVID one using biomarkers extracted from CT images: i.e. a binary classification problem.

The Python language was used also for the development of artificial intelligence models since many libraries, as tensorflow, keras, scikit-learn and hyperopt, were useful for the purposes of this work.

Four classifiers have been implemented; two called "*radiomic models*" and two called "*clinical models*".

The radiomic models implemented are:

• *radiomic model first order*, using only the first order features;

• *radiomic model all*, built with the use of all the extracted features, therefore first order, GLCM and GLSZM.

The number of patients used for the radiomic models is 1028 (646 COVID, 382 no-COVID).

The clinical models implemented are:

• *clinical model bilateral lung*, built with only features relating to the bilateral lung. The number of patients used is 965(616 COVID, 349 no-COVID);

• *clinical model all*, built with features related to bilateral lung and sub-regions. 863 patients were used (550 COVID, 313 no-COVID).

The number of patients for clinical models is reduced compared to that of radiomic models due to the non-applicability of the Gaussian model in all patients analyzed; the features have been extracted with QCT.

For each classifier, since classification is a supervised learning problem, an MLP neural network has been implemented.

The construction pipeline was the same for all classifiers and below is presented the workflow with which they were built:
- a LASSO regression was performed to select the features;
- a simple classifier has been created; an early stopping algorithm was then added to avoid overfitting;
- an hyperparatmeters tuning has been done to find the best parameters for each classifier. Also in this case an early stopping algorithm was ,added to avoid overfitting;
- A K-Fold Cross-Validation was performed on the best model obtained with hyper parametrization.

The performance of each classifier was assessed calculating accuracy, confusion matrix, ROC curve and AUC.

## 3. Results

### 3.1. QCT

The construction of a single script for the extraction of biomarkers from bilateral lung and sub-regions allowed to extend the Gaussian model in regions where it was not applicable. The figure 2 shows an example of a histogram of the distribution of the HU of the bilateral lung. The presence of the Gaussian fit and therefore the distinction between ventilated and non-ventilated lung can be observed.

The percentage of patients with non-applicability of the model in bilateral lung prior to extension was 15%; with the extension it was reduced to 6%.

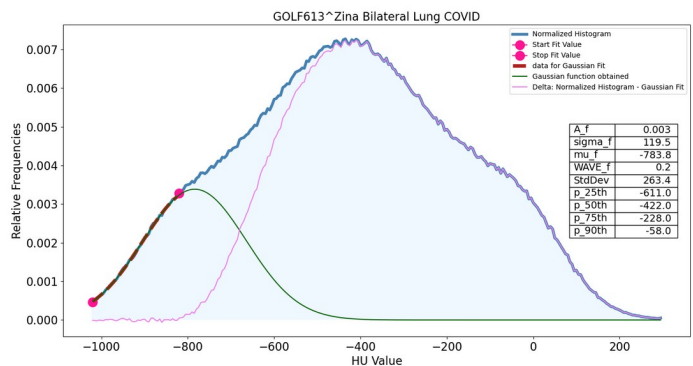

Figure 2: Example of Gaussian model extension of an HU distribution

### 3.2. Classifiers

The results of the models' performances are reported in table1, together with the number of patients and the number of relevant features used to build the model.

The resulting accuracy of the models were in range 0.76-0.87, as the integral of the receiver operating characteristic curve (AUC), indicating overall good performances in the diagnostic task. The two classifiers with a greater number of features (*clinical model all* and *radiomic model all*) resulted as the models with the highest accuracy; geometrical information and information in the image textures were different in the two population and so helpful for COVID- 19 classification. The ROC curves for train and test sets of each models are reported in figure 3.

The best diagnostic performances were associated to the *radiomic model all*, based on 25 features selected with LASSO optimization from a set of 139 different metrics, after a 4-fold cross-validation showed AUC=0.87 ± 0.02 and accuracy = 0.81 ± 0.02. Confusion matrices of train and test sets of this model are reported in figure 4

|  | Radiomic 1st | Radiomic All | Clinica Bilateral | Clinical all |
|---|---|---|---|---|
| Number of patients | 1028 | 1028 | 965 | 866 |
| Number of initial features | 19 | 129 | 28 | 124 |
| Number of relevant features | 9 | 21 | 11 | 25 |
| Train accuracy | 0.715 | 0.863 | 0.761 | 0.816 |
| Test accuracy | 0.708 | 0.807 | 0.712 | 0.788 |

| | Train AUC | 0.777 | 0.929 | 0.822 | 0.865 |
| --- | --- | --- | --- | --- | --- |
| | Test AUC | 0.767 | 0.874 | 0.759 | 0.841 |

Table 1: the performances and main characteristic of each model

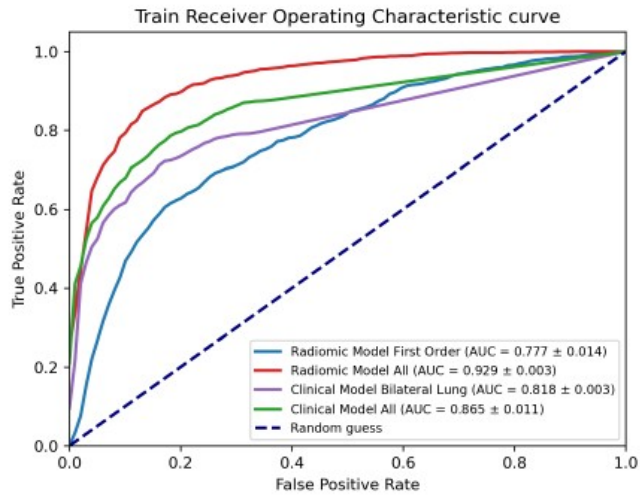

(a) Train ROC curve for every classifier.

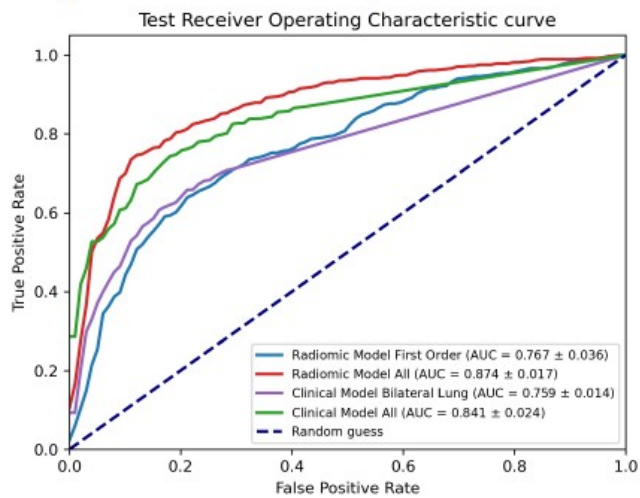

(b) Test ROC curve for every classifier.

Figure 3 : Training and test ROC curve obtained for every classifiers. For each model the mean ROC curve obtained with k-fold cross-validation is represented. In legend you will find the average AUC values

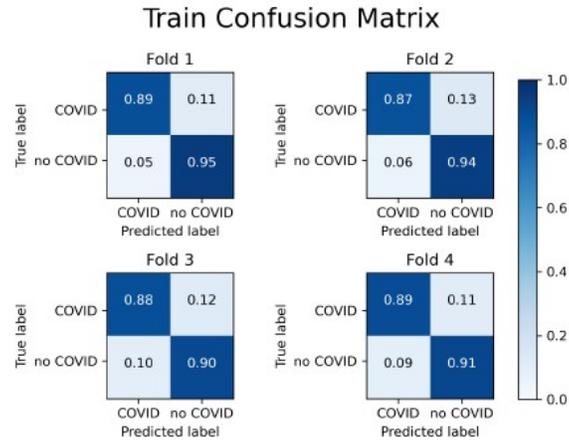

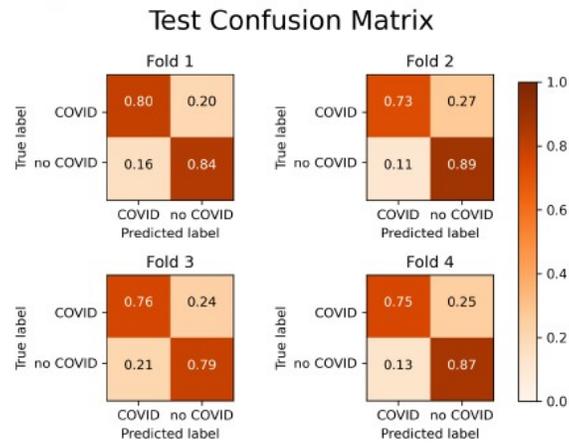

Fig 4 *Radiomic model all* training and test confusion matrix for K-Fold Cross-Validation.

## 4. Discussion

The first aim of this work was to develop a pipeline for automatic extraction of some biomarkers from CT images. Two different approaches were adopted: the former, following Niguarda's team previous works, was focused on histogram-based metrics describing well-aerated and pathological tissues of the lung while the latter was an application of already described metrics called radiomic features derived from image textures analysis.

Particular attention was dedicated to extend the applicability of the clinical model based on the Gaussian fit of histogram to a higher number of patients using information of lung subregions. The results obtained for the Gaussian model show a notable improvement in its application. Simultaneous analysis of the five regions made it possible to improve the quantitative analysis by allowing the extraction of biomarkers in patients and/or regions where the Gaussian model was not applicable due to intrinsic limitations of the fitting of histogram data.

Following the developments implemented in this work, the applicability of the Gaussian model increased from 85% to 94% of the CT images analyzed. Furthermore, the extraction of biomarkers in the sub-regions, which provide information on the location and severity of the disease, was found to be relevant for the COVID-19 classifier.

The advantage of using Gaussian model and the related metrics (WAVE and biomarkers from histogram of the "non-healthy" tissue) is that you have control over the analysis and extraction, it is a highly interpretable model; on the other hand, the features extracted with PyRadiomics are more specific but less interpretable. However, Gaussian model have some limitations. WAVE results are not clinically validated with independent physiological measurements. Furthermore, in some patients with severe and diffuse lung opacities or solidifications the Gaussian fitting of CT histograms data is impossible and the model is not applicable. By contrast, the extraction of features with PyRadiomics is applicable to all patients and is also a universal model being an opensource package; the latter was in fact also used by other working groups in the analysis of chest CT images of COVID-19 patients.

All classifiers showed a good ability to classify COVID and non-COVID populations. The resulting accuracy of the models were in range 0.76-0.87, as the integral of the receiver operating characteristic curve (AUC), indicating overall good performances in the diagnostic task. The first order radiomic model presented the worst results, mainly observed in CM obtained with k-fold cross-validation. The results obtained through the k-Fold Cross-Validation are considered the most reliable because it is a technique that allows to obtain less optimistic and more reliable estimates.

The mean AUC values and the mean accuracy of each model are then used for comparison. In fact, K-Fold Cross-Validation is commonly used to compare and select a model for a given predictive modeling problem. In this case, by making this comparison, the best classifier can be selected.

The two classifiers with a greater number of features (the *clinical model all* in which both bilateral and Sub-ROI biomarkers have been used and the *radiomic model all* in which first order, GLCM and GLSZM features were used) resulted as the models with the highest AUC:

• radiomic model: 87% vs 78%;

• clinical model: 85% vs 76%.

This mean that geometrical information of the sub-regions and the information in the image textures described by radiomic features were different in the two population and so helpful for COVID-19 classification.

The best diagnostic performances were associated to the *radiomic model all*, based on 25 useful/independent features selected with LASSO optimization from a set of 139 different metrics, after a 4-fold cross-validation showed AUC=0.87 ± 0.02 and accuracy = 0.81 ± 0.02.

Despite the good results obtained, this work had some limitations. As already discussed in the previous section, the Gaussian model is not always applicable to all patients. Furthermore, in this work all CT images came from a single institution: this exclude many sources of variation due to RX devices, protocols parameters and patients selection. Another possible bias was due to the lack of "external" test even if k-fold validation was used. Finally, the number of cases (around 103) and the imbalance between the two categories could be considered sub-optimal for robust modeling.